# Inverse magnetocaloric effect in ferromagnetic Ni-Mn-Sn alloys

Thorsten Krenke[1], Eyüp Duman[1], Mehmet Acet[1], Eberhard F. Wassermann[1], Xavier Moya[2], Lluis Mañosa[2] and Antoni Planes[2]

[1] *Fachbereich Physik, Experimentalphysik, Universität Duisburg-Essen, D-47048 Duisburg, Germany*

[2] *Departament d'Estructura i Constituents de la Matèria, Facultat de Física, Universitat de Barcelona, Diagonal 647, E-08028 Barcelona, Catalonia, Spain*

'These authors contributed equally to this work'

**The magnetocaloric effect (MCE) in paramagnetic materials has been widely used for attaining very low temperatures by applying a magnetic field isothermally and removing it adiabatically. The effect can be exploited also for room temperature refrigeration by using recently discovered giant MCE materials.[1-3] In this letter, we report on an inverse situation in Ni-Mn-Sn alloys, whereby applying a magnetic field adiabatically, rather than removing it, causes the sample to cool. This has been known to occur in some intermetallic compounds, for which a moderate entropy *increase* can be induced when a field is applied, thus giving rise to an inverse magnetocaloric effect.[4,5] However, the entropy change found for some ferromagnetic Ni-Mn-Sn alloys is just as large as that reported for giant MCE materials, but with opposite sign. The giant inverse MCE has its origin in a martensitic phase transformation that modifies the magnetic exchange interactions due to the change in the lattice parameters.**

The magnetocaloric effect (MCE) is a temperature change that occurs when a magnetic field is applied under adiabatic conditions. In general, an isothermal application of a magnetic field decreases the configurational entropy of the spin structure. A subsequent adiabatic demagnetisation produces a spin re-disordering by the thermal energy provided by the phonon bath of the isolated sample. This causes cooling.

While cooling by adiabatic demagnetisation has been restricted to low temperature cryogenic applications for many decades, the recent discovery of the giant magnetocaloric effect around room temperature[1] has opened up the possibility to exploit this effect for room temperature refrigeration as an environment-friendly alternative to conventional vapour-cycle refrigeration. This has prompted intensive research in this field.[2-4,6] MCE is largest in the immediate vicinity of magnetic transition temperatures, where the magnetic ordering configuration changes rapidly with varying temperature. The giant MCE found in $Gd_5(Si_xGe_{1-x})_4$ alloys is due to concurrently occurring first order structural and magnetic phase transitions, whereby the transition spreads over a narrow temperature range of about 10 K. The reported entropy change $\Delta S$ at room temperature is about $-20$ J K$^{-1}$ kg$^{-1}$.[2] By varying the composition, the transition temperatures and, thereby, the MCE working point can be changed in a broad temperature range.[7]

As opposed to cooling by adiabatic demagnetisation, cooling by adiabatic magnetisation (inverse MCE) requires an increase of configurational entropy on applying a magnetic field. The inverse MCE is observed in systems where first order magnetic transformations, such as antiferromagnetic/ferromagnetic (AF/FM), AF-collinear/AF-non-collinear, AF/ferrimagnetic (FI), etc., take place. This leads to the occurrence of magnetically inhomogeneous states in the vicinity of the magnetic transformation temperature. Some examples of such systems are $Fe_{0.49}Rh_{0.51}$ (AF/FM),[8]

Mn$_5$Si$_3$ (AF-collinear/ AF-non-collinear AF),[4] Mn$_{1.96}$Cr$_{0.05}$Sb (AF to FI),[4] and Mn$_{1.82}$V$_{0.18}$Sb (AF-FI).[5] Because of the presence of mixed magnetic exchange interactions, it is thought that the application of an external magnetic field leads to further spin disorder in these systems giving rise to an increase of the configurational entropy.[4,9]

An interesting group of alloys in relation to the MCE are ferromagnetic Ni-Mn based Heusler alloys, of which Ni-Mn-Ga is the most studied series (also because of the presence of the magnetic shape memory effect). In a broad composition range, these alloys transform from a parent austenitic state to a product martensitic state. In each of the states, the magnetic coupling is ferromagnetic, however, with different ferromagnetic exchange. The martensitic transformation in Ni-Mn-Ga is first order with a narrow thermal hysteresis, and the application of a magnetic field in the vicinity of this transition leads to a large MCE.[9-12] It has been shown that the maximum MCE is obtained when structural and magnetic transition temperatures lie close to each other.[9] For samples with compositions close to Ni$_2$MnGa stoichiometry, an inverse MCE has been reported. This effect, however, vanishes as the magnetic field increases, and the standard MCE is observed at high fields. Such an inverse MCE is an extrinsic effect arising from the coupling at the mesoscale between the martensitic and magnetic domains.[11]

Heusler-type Ni$_{0.50}$Mn$_{0.50-x}$Sn$_x$ alloys, the properties of which we discuss in this letter, exhibit structural and magnetic phase transformations.[13] This is found in alloys with compositions lying in the narrow range $0.13 \leq x \leq 0.15$, for which an inverse MCE is observed that is at least three times larger than in the compounds mentioned above. In this work, we examine the details of the magnetic field induced entropy changes at temperatures in the vicinity of the transformations of two samples having compositions corresponding to the limits of this range. We study the magnetocaloric properties through the temperature and field dependences of the magnetisation $M(T)$ and $M(H)$.

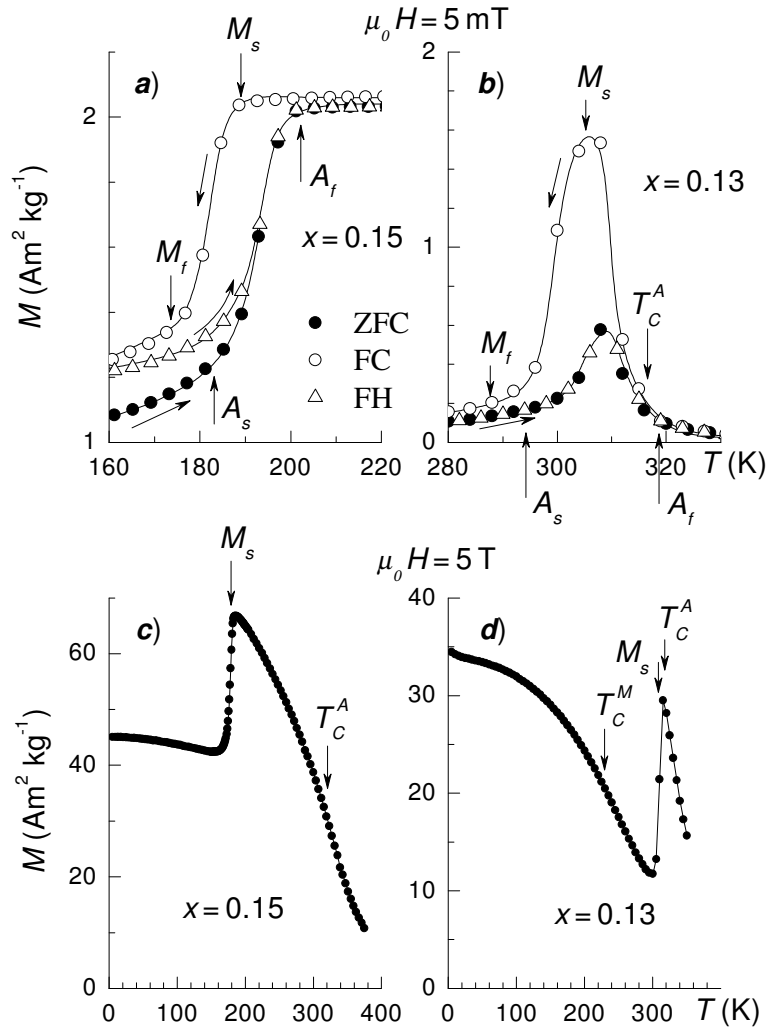

**Figure 1 Temperature dependence of the magnetization of $Ni_{0.50}Mn_{0.50-x}Sn_x$ alloy in low and high external magnetic fields.** FC, field-cooled; ZFC, zero-field-cooled; FH, field-heated. **a,b,** External magnetic field 5 m, for $x = 0.15$ and $x = 0.13$ respectively. **c,d,** External magnetic field of 5 T, for $x = 0.15$ and $x = 0.13$ respectively.

Fig. 1a and 1b show $M(T)$ in the vicinity of the martensitic transformation for $x = 0.15$ and $x = 0.13$ respectively. The measurements are made in a low external magnetic field of 5 mT. The data in Fig.1c and 1d show $M(T)$ in the temperature range $5\ K \leq T \leq 400\ K$ and is taken in 5 T. Both samples are FM in the austenitic state below their Curie temperatures $T_C^A$.

In low external magnetic fields, a splitting between the zero-field-cooled (ZFC) and the field cooled (FC) data below a ferromagnetic Curie temperature indicates to the presence of magnetically inhomogeneous states,[14] whereas a difference in the FC and field-heated (FH) data is due to the thermal hysteresis of the structural transformation. The values of the austenite and martensite phase Curie temperatures $T_C^A$ and $T_C^M$, as well as the martensite "start" and "finish" temperatures $M_s$ and $M_f$ shown in Fig. 1 are taken from the results of earlier magnetisation and calorimetric measurements.[13] The width of the hysteresis for both samples were also determined from the calorimetric measurements to be about 20 K. For both samples, the splitting in the FC and ZFC magnetization data around the martensitic transformation in low external magnetic fields indicates the development of non-FM components leading to the occurrence of a magnetically inhomogeneous state. The drop in $M(T)$ below $M_s$ suggests that these components are likely to be AF in nature.

The decrease in $M(T)$ below $M_s$ is still observed in the data taken in a high external field of 5 T for both samples. The features in the transition neither smear out nor does the change in the magnetization vanish or change sign as in Ni-Mn-Ga. The $x = 0.15$ sample undergoes a martensitic transformation at $M_s = 189$ K, and both the austenite and the martensite phases are essentially FM with $M(T)$ running at lower values in the martensite phase down to the lowest temperatures (Fig. 1c). For this sample, $T_C^A = 320$ K. For the alloy with $x = 0.13$, the martensitic transition begins at $M_s = 307$ K, which lies just below the Curie temperature $T_C^A = 311$ K (Fig. 1d). For this sample, $M_s$ lies very close to $T_C^A$, and a Curie temperature of the martensitic state is observed at $T_C^M = 230$ K. In this sample, there is no long-range magnetic ordering in the temperature range $T_C^M \leq T \leq M_s$, where the martensitic transformation progresses from an austenitic phase with long range FM order to a martensitic phase with non-FM components. This gives rise to a magnetically inhomogeneous state. Long-range ordering sets in once again at $T_C^M$. The presence of the non-FM components in the range $T_C^M \leq T \leq M_s$ is evidenced from the non-saturating feature of the $M(H)$ curves, which is discussed below in relation to Fig. 3a.

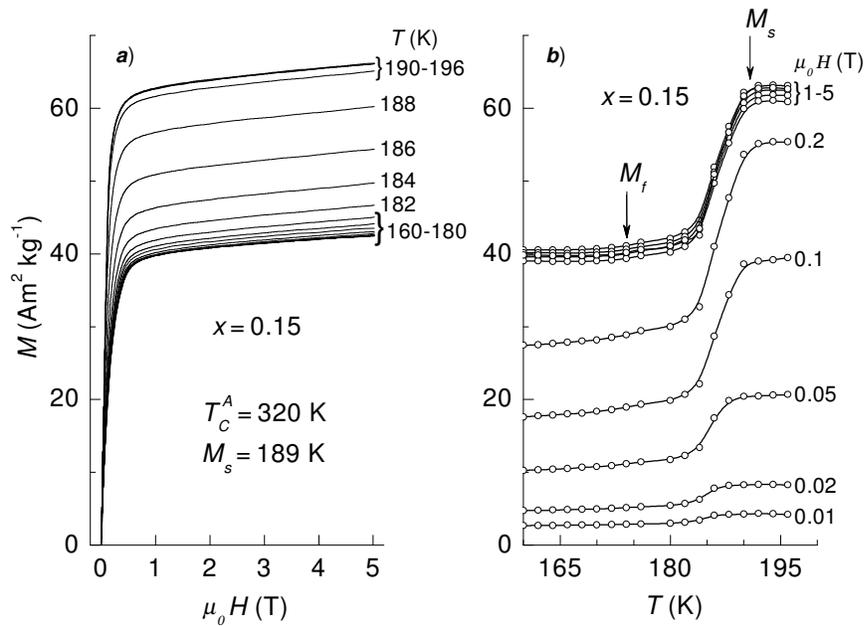

**Figure 2 Magnetic field dependence of the magnetization near the martensitic transformation for *x* = 0.15. a**, Curves of *M(H)* in the temperature range 160 K ≤ *T* ≤ 196 K.. **b**, *M(T)* for *x* = 0.15 obtained from the *M(H)* data in **a**.

Fig. 2a shows the field dependence of the magnetisation $M(H)$ for the sample with $x = 0.15$ measured in the range 160 K $\leq T \leq$ 196 K in 2 K steps. In the temperature intervals 160 K-180 K and 190 K-196 K, where $M(H)$ is weakly temperature dependent, the data are labelled together. $M(H)$ do not saturate, but acquire a substantial high field susceptibility due to the inhomogeneous nature of the magnetic state. The temperature dependence of the magnetisation $M(T)$ at constant magnetic fields obtained from these data is plotted in Fig. 2b. For all magnetic fields, $M(T)$ features a step-like behaviour below $M_s$ with the step height $\Delta M = M_A - M_M$, where $M_A$ and $M_M$ are the magnetisations of the upper (austenitic) and lower (martensitic) limits of the step. $\Delta M$ is positive for all external fields. It initially increases with increasing field and eventually remains constant.

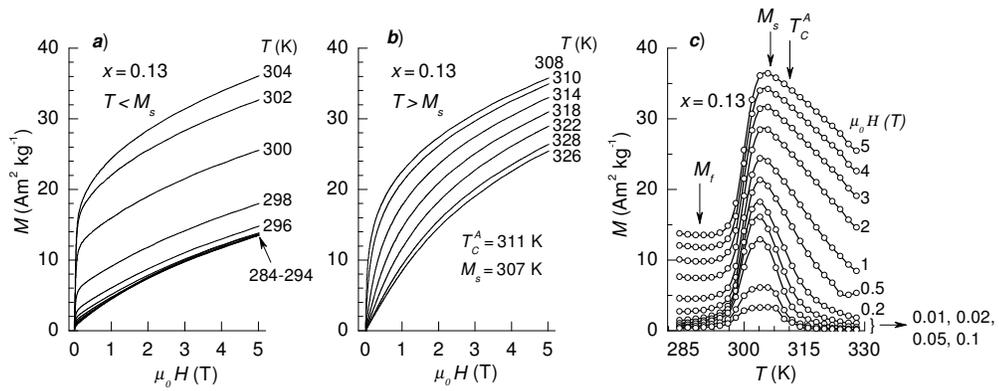

**Figure 3 Magnetic field dependence of the magnetization near the martensitic transformation for $x$ = 0.13.** Curves of $M(H)$ curves in the temperature ranges **a**, 284 K ≤ $T$ ≤ 304 K and **b**, 308 K ≤ $T$ ≤ 326 K. **c**, Curves of $M(T)$ for $x$ = 0.13 obtained from the $M(H)$ data of parts **a** and **b**.

Fig. 3a and 3b show $M(H)$ for $x$ = 0.13 plotted for $T < M_s$ and $T > M_s$ respectively. The magnetisation for $T < M_s$ steadily drops with decreasing temperature as the martensitic transformation progresses, whereas for $T > M_s$, it increases with decreasing temperature down to about $T_C^A$. In the studied temperature range above $T_C^A$, the presence of FM short range correlations in the paramagnetic state causes $M(H)$ to deviate from linearity and bend. On the other hand, *M(H)* also bends for $M_f \leq T \leq M_s$, where the austenite and martensite phases are mixed. This is only possible if non-FM components are present in the FM matrix. The exact nature of the non-FM components in this material is not known, but it is likely to be in the form of short range antiferromagnetic correlations. The antiferromagnetism would be caused by shortened Mn-Mn bonds due to the occurrence of Mn-Mn nearest neighbors. Therefore, for $M_f \leq T \leq M_s$, the ferromagnetic component (remaining austenite) governs the steep nature of the initial slope of *M(H)*, whereas the non-ferromagnetic component keeps *M(H)* from saturating. *M(T)* curves constructed from these data are plotted in 3c in the vicinity of the magnetic and structural transition temperatures. A step-like behaviour as in the data of the *x* = 0.15 sample is also found here between $M_f$ and $M_s$.

We estimate the field induced entropy change Δ*S* around the martensitic transformation for both samples using the relationship[15]

$$\Delta S(T, H) = \mu_0 \int_0^H \left( \frac{\partial M}{\partial T} \right)_H dH,$$

where $\mu_0$ is the permeability of free space. The sign of Δ*S* is determined by the sign of ∂*M*/∂*T*. In the present case, the sign is positive for both samples in the range $M_f \leq T \leq M_s$ as seen in the plots in Fig. 4. The maxima in Δ*S* vs. *T* in a magnetic field of 5 T correspond to about 15 J K$^{-1}$ kg$^{-1}$ for *x* = 0.15 and 18 J K$^{-1}$ kg$^{-1}$ for *x* = 0.13. For *x* = 0.13, Δ*S* is negative for $T > T_C^A$, since ∂*M*/∂*T* < 0 for this case. The values of Δ*S* are smaller in this temperature range because the magnetic transition is second order and *M* changes over a broader temperature range than that for the first order martensitic transformation. The values for Δ*S* at $\mu_0 H$ = 5 T for the present samples are comparable to the value observed for $Gd_5Si_2Ge_2$. However, in this material, the value is negative with Δ*S* ≈ −19 J K$^{-1}$ kg$^{-1}$.[2]

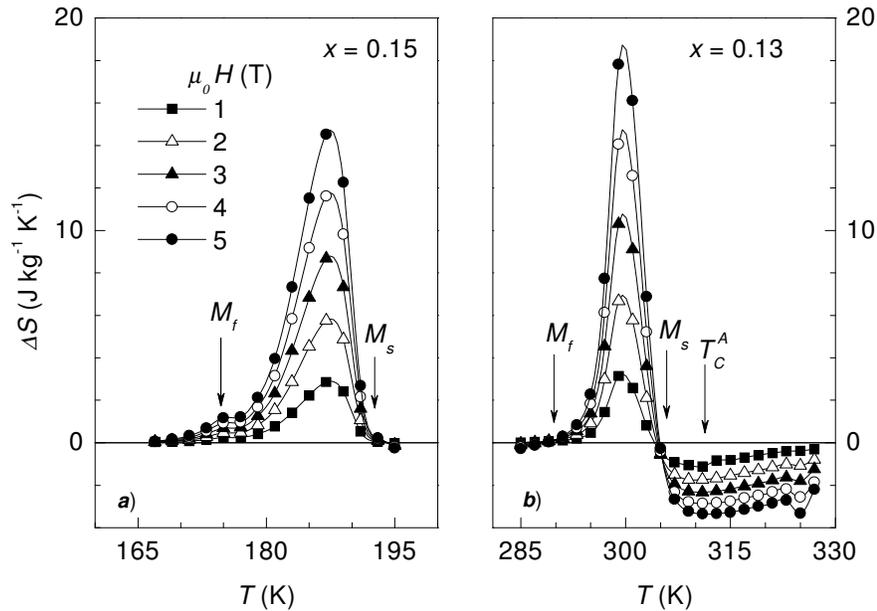

**Figure 4 Entropy change $\Delta S$ for $x = 0.15$ and $x = 0.13$.** The value of $\Delta S$ is positive for both compositions in the temperature range $M_f < T < M_s$.

The Ni-Mn-Sn system stands out as having a sizable *positive* magnetic entropy change ($\Delta S > 0$) below $M_s$. This effect is associated with the martensitic transition that undergoes in this alloy which causes the magnetization to decrease. Although a microscopic interpretation of these results would require *ab initio* calculations in non-stoichiometric systems, results for stoichiometric alloys show that the magnetic interactions in $Ni_2MnSn$ are very sensitive to the atomic distances.[16] Therefore the change in the atomic distance at the martensitic transition modifies the magnetic properties of the alloy.

For $Mn_{1.96}Cr_{0.05}Sb$, an inverse MCE of the same magnitude as in the present $x = 0.13$ sample at a field of 2 T has been reported.[4] However, in $Mn_{1.96}Cr_{0.05}Sb$ the MCE does not increase any further beyond 7 J $K^{-1}$ $kg^{-1}$ with increasing magnetic field, whereas in the $x = 0.13$ sample, values up to 20 J $K^{-1}$ $kg^{-1}$ are attained at 5 T, which is also not the maximum field limit. This means that the inverse MCE could attain even higher values in higher applied magnetic fields. However, even at 5 T, the magnitude of the effect is comparable to that in giant magnetocaloric materials, and it is the largest attained so far for the inverse MCE. It is also worth noting that in $Mn_{1.96}Cr_{0.05}Sb$ the transition at high fields is smeared out over a broad temperature interval while in Ni-Mn-Sn, the transition is hardly affected by the magnetic field.

It has been suggested that a possible application of the inverse MCE other than cooling *per se*, would be to use it as a heat-sink for heat generated when conventional MCE material is magnetized prior to cooling by adiabatic demagnetisation.[17] Therefore, the inverse MCE opens up the possibility to increase room temperature refrigeration efficiency by employing materials showing this effect in composites with conventional MCE material.

**Methods**

Ingots were prepared by arc melting the pure metals under argon atmosphere in a water cooled Cu crucible, and their compositions were determined by energy dispersive x-ray photoluminescence analysis (EDAX) to correspond to $x = 0.13$ and $x = 0.15$. The samples were then encapsulated under argon in quartz glass and annealed at 1273 K for two hours and subsequently quenched in an ice-water mixture. Magnetisation measurements were carried out using a superconducting quantum interference device magnetometer in fields up to 5 T.

1. Pecharsky, V.K. & Gschneidner Jr., K.A. Giant magnetocaloric effect in $Gd_5(Si_2Ge_2)$. *Phys. Rev. Lett.* **78**, 4494-4497 (1997).

2. Pecharsky, V.K. & Gschneidner Jr., K.A. $Gd_5(Si_xGe_{1-x})_4$: an extremum material. *Adv. Mater.* **13**, 683-686 (2001).

3. Tegus, O., Brück, E., Buschow, K.H.J. & de Boer, F.R. Transition-metal-based magnetic refrigerants for room-temperature applications. *Nature*, **415**, 150-152 (2002).

4. Tegus, O., Brück, E., Zhang, L., Dagula, Buschow, K.H.J., de Boer, F.R. Magnetic phase transitions and magnetocaloric effects. *Physica B* **319**, 174-192 (2002).

5. Zhang, Y. Q. & Zhang, Z. D. Giant magnetoresistance and magnetocaloric effects of the $Mn_{1.82}V_{0.18}Sb$ compound. *J. Alloys and Comp.* **365**, 35-38 (2004).

6. Provenzano, V., Shapiro, A.J. & Shull R.D. Reduction of hysteresis losses in the magnetic refrigerant $Gd_5Si_2Ge_2$ by addition of iron. *Nature* **429**, 853-857 (2004).

7. Pecharsky, V.K. & Gschneidner Jr., K.A. Tunable magnetic regenerator alloys with a giant magnetocaloric effect for magnetic refrigeration from ~20 to ~290 K. *Appl. Phys. Lett.* **70**, 3299-3301 (1997).

8. Nikitin, S. A., Myalikgulyev, G., Tishin, A. M., Annaorazov, M. P., Asatryan, K. A. & Tyurin, A. L. The magnetocaloric effect in $Fe_{49}Rh_{51}$ compound. *Physics Letters A* **148**, 363-366 (1990).

9. Pareti, L., Solzi, M., Albertini, F. & Paoluzi, A. Giant entropy change at the co-occurrence of structural and magnetic transitions in the $Ni_{2.19}Mn_{0.81}Ga$ Heusler alloy. *Eur. Phys. J. B* **32**, 303-307 (2003).

10. Marcos, J., Planes, A., Mañosa, Ll., Casanova, F., Batlle X., Labarta, A. & Martínez, B. Magnetic field induced entropy change and magnetoelasticity in Ni-Mn-Ga alloys. *Phys. Rev. B* **66**, 224413 (2002).

11. Marcos, J., Mañosa, Ll., Planes, A., Casanova, F., Batlle, X. & Labarta, A. Multiscale origin of the magnetocaloric effect in Ni-Mn-Ga shape-memory alloys. *Phys. Rev. B* **68**, 094401 (2003).

12. Hu, F., Shen, B. & Sun, J. Magnetic entropy change in $Ni_{51.5}Mn_{22.7}Ga_{25.8}$ alloy. *Appl. Phys. Lett.* **76**, 3460-3462 (2000).

13. Krenke, T., Acet, M., Wassermann, E.F., Moya, X., Mañosa, Ll. & Planes, A. Martensitic transitions and the nature of ferromagnetism in the austenitic and martensitic states of Ni-Mn-Sn alloys. Submitted to *Phys. Rev. B.*

14. Duman, E., Acet, M., Elerman, Y., Elmali A. & Wassermann, E. F. Magnetic interactions in $Pr_{1-x}Tb_xMn_2Ge_2$. *J. Magn. Magn. Mater.* **238**, 11-21 (2002).

15. Tishin, A.M. & Spichkin, Y.I. *The magnetocaloric Effect and its Applications,* Institute of Physics Publishing, Bristol 2003.

16. Sasioglu E., Sandratskii L.M. & Bruno P. *Phys. Rev. B* **70**, 024427 (2004).

17. Joenk, R.J. Adiabatic magnetisation of antiferromagnets. *J. Appl. Phys.* **34**, 1097-1098 (1963).

Acknowledgements This work was supported by Deutsche Forschungsgemeinschaft (GK277) and CICyT (Spain), project MAT2004-1291. XM acknowledges support from DGICyT (Spain).

Competing Interests statement The authors declare that they have no competing financial interests.

Correspondence and requests for materials should be addressed to M.A. (e-mail: macet@agfarle.uni-duisburg.de).